\documentclass[aps,pra,reprint,twocolumn,superscriptaddress,amsmath,longbibliography,nofootinbib]{revtex4-2}
\usepackage{amsmath}
\usepackage{amsfonts}
\usepackage{amssymb}
\usepackage{braket}
\usepackage{tcolorbox}
\usepackage{xcolor}
\usepackage{graphicx}
\usepackage{enumerate}
\usepackage{appendix}

\usepackage[colorlinks=true,
            linkcolor=blue,
            citecolor=blue,
            urlcolor=blue]{hyperref}

\begin{document}

\title{Adiabatic reverse annealing is robust to low-temperature decoherence}

\author{An Le} 
\email{lean2@msu.edu}
\affiliation{Department of Physics and Astronomy, Michigan State University, East Lansing, Michigan 48824, USA}
\date{\today}

\author{Christopher Baldwin}
\email{baldw292@msu.edu}
\affiliation{Department of Physics and Astronomy, Michigan State University, East Lansing, Michigan 48824, USA}
\date{\today}

\begin{abstract}
Adiabatic reverse annealing (ARA) is an improvement to conventional quantum annealing (QA) that uses an initial guess at the desired ground state to circumvent problematic phase transitions.
Despite encouraging results in the closed-system setting, Ref.~\cite{standard_QA_outperform_ARA} has suggested on the basis of numerical simulations that ARA may lose its advantage in the presence of decoherence.
Here, we revisit this problem from a more analytical perspective.
Using the $p$-spin model as a solvable example, together with the adiabatic master equation to describe the effects of the environment (valid at weak coupling), we show that ARA can in fact succeed in open systems but that the temperature of the environment plays a key role.
We first demonstrate that, in the adiabatic limit, the system will follow the instantaneous equilibrium state as long as the protocol does not pass through any (finite-temperature) phase transitions.
Given this, there are two distinct mechanisms by which ARA can break down at high temperature: either there are no paths that avoid transitions, or the equilibrium state itself is disordered.
When the temperature is sufficiently low that neither of these occur, then ARA succeeds.
Remarkably, there are even situations in which the environment benefits ARA: we find parameter values for which no transition-avoiding paths exist at zero temperature but such paths appear at non-zero temperature.
\end{abstract}

\maketitle

\section{Introduction} \label{sec:introduction}

Quantum annealing (QA) is a branch of adiabatic quantum computation that aims to solve hard optimization and spin-glass problems by exploiting quantum fluctuations~\cite{finnila1994quantum, QA_transverse_field, brooke1999quantum, farhi2001quantum, santoro2002theory, santoro2006optimization, morita2008mathematical, johnson2011quantum, boixo2014evidence, PhysRevX.4.021041, kumar2018quantum, jiang2018quantum, RevModPhys.90.015002, hauke2020perspectives, rajak2023quantum, ding2024effective}.
Such problems can often be formulated as finding the ground state of an Ising Hamiltonian $H_0$ involving $N$ spin-$1/2$s, usually diagonal in the $\hat{\sigma}^z$ (``computational'') basis.
In conventional QA, an additional transverse field $H_{\textrm{TF}}$ is applied, so that the system evolves under the Hamiltonian
\begin{equation} \label{eq:conventional QA}
H_S(s) = s H_0 + (1-s) H_{\textrm{TF}},
\end{equation}
with $s \in [0, 1]$ controlling the relative strength of the two terms.
The spins are prepared in the ground state of $H_{\textrm{TF}}$, and $s$ is increased from 0 to 1.
If this variation is sufficiently slow, the adiabatic theorem of quantum mechanics~\cite{kato1950adiabatic, Messiah1962, jansen2007bounds} guarantees that the spins will remain in their instantaneous ground state throughout the protocol, which is ultimately the desired ground state of $H_0$.
Although conceptually simple, a large body of evidence by now suggests that conventional QA will require prohibitively long annealing times to find the ground states of challenging optimization problems, at least generically~\cite{simple_glass_models_QA,altshuler2010anderson,Jorg2010FirstOrder,Young2010First,Hen2011Exponential,Farhi2012Performance,Bapst2013Quantum,Knysh2016ZeroTemperature,Baldwin2018Quantum}.
In physical terms, the failure of conventional QA is due to discontinuous phase transitions in the ground state of $H_S(s)$, which are themselves a consequence of the ``rugged energy landscape'' in $H_0$.

With this in mind, a number of works have turned to more elaborate annealing protocols than Eq.~\eqref{eq:conventional QA}, hoping to find mechanisms for suppressing or circumventing the problematic phase transitions.
Prominent examples include non-stoquastic annealing~\cite{Seki2012Quantum,Seki2015Quantum,Hormozi2017Nonstoquastic,Albash2019Role}, inhomogeneous annealing~\cite{Susa2018Exponential,Susa2018Quantum,Adame2020Inhomogeneous,Anomalously_slow_IQA}, and iterated reverse annealing~\cite{Marshall2019Power,Passarelli2020Reverse,Rocutto2021Quantum,Bando2022Breakdown,Mehta2025Unraveling}.

Here we focus on the variant known as ``adiabatic reverse annealing'' (ARA)~\cite{PerdomoOrtiz2011Study,Chancellor2017Modernizing}.
In ARA, an additional longitudinal field is added to the Hamiltonian:
\begin{equation} \label{eq:ARA Hamiltonian}
H_S(s,\lambda) = s H_0 + (1-s) \lambda H_{\textrm{TF}} + (1-s)(1-\lambda) H_{\textrm{LF}},
\end{equation}
where
\begin{equation} \label{eq:longitudinal fields}
H_{\textrm{TF}} = -\sum_{j=1}^N \hat{\sigma}_j^x, \qquad H_{\textrm{LF}} = -\sum_{j=1}^N \epsilon_j \hat{\sigma}_j^z.
\end{equation}
Each $\epsilon_j$ takes values in $\{1, -1\}$, and the configuration $\{ \epsilon_j \}$ is meant to be an initial guess at the ground state of $H_0$.
The relative strength of the longitudinal and transverse fields is controlled by the new parameter $\lambda$, with $\lambda = 1$ reducing to conventional QA and $\lambda = 0$ being a purely classical Ising model.
As for conventional QA, we vary $s$ and $\lambda$ in time (often writing $H_S(t)$ as shorthand for $H_S(s(t), \lambda(t))$), specifically increasing $(s, \lambda)$ from $(0, 0)$ to $(1, 1)$.
The ground state at the beginning of the protocol is the (known) configuration $\{ \epsilon_j \}$, and the ground state at the end is that of $H_0$.
ARA is efficient if there are paths from $(0, 0)$ to $(1, 1)$ which do not cross discontinuous phase transitions.

Studies of ARA have yielded a number of encouraging results~\cite{reverse_annealing_p_spin_model,Yamashiro2019Dynamics,Arai2021Mean,Arai2022MeanField,Passarelli2023Counterdiabatic}, including a recent experimental investigation~\cite{Zhang2024Cyclic}.
In particular, Ref.~\cite{reverse_annealing_p_spin_model} has analytically shown that ARA indeed allows one to circumvent phase transitions in the solvable family of mean-field models known as the $p$-spin model (defined below).
However, the recent Ref.~\cite{standard_QA_outperform_ARA} has numerically studied how noise from an environment --- as all near-term quantum computers are subject to --- affects ARA in the same model.
The overall conclusion is that ARA largely loses its advantage compared to conventional QA, at least for the parameter values considered in Ref.~\cite{standard_QA_outperform_ARA}.

Given this potentially discouraging result, here we conduct a complementary investigation of open-system ARA (in the $p$-spin model) via analytical means, which allows us to avoid complications arising from finite $N$.
One minor drawback is that, strictly speaking, our results are limited to \textit{approximate} optimization: whether open-system ARA can produce states at energies within a finite-percentage threshold of the optimum.
Yet in return, we can draw unambiguous conclusions.
We find that ARA can in fact succeed in the presence of an environment, although the temperature of the environment must be sufficiently low.
Remarkably, a low but non-zero temperature can even produce exponential \textit{improvement} in ARA, allowing it to circumvent phase transitions in some situations where it may not have been able to at zero temperature.

\begin{figure}[t]
\begin{center}
\includegraphics[width= 1.\linewidth, trim = 10 0 10 10, clip]{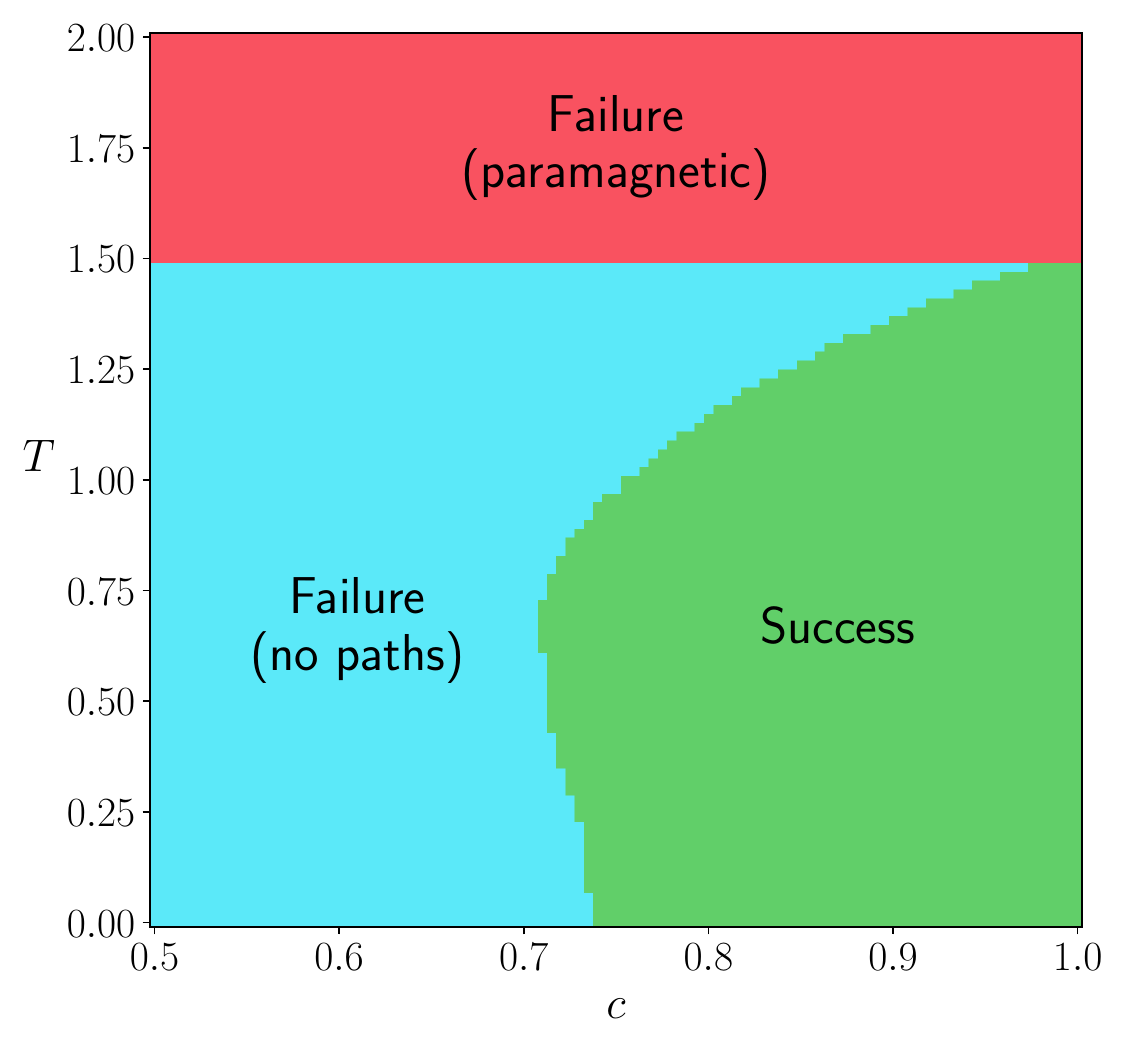}
\caption{Success/failure of open-system ARA in the $p$-spin model ($p = 3$), as a function of the environment temperature $T$ and the fraction $c$ of spins which have the same orientation in the initial guess as in the ground state. Note that there are two failure modes for ARA: there can be no paths from $(s, \lambda) = (0, 0)$ to $(1, 1)$ which do not cross phase transitions (``no paths'', shown in blue); or the equilibrium state at the end of the protocol can be in the paramagnetic phase of the model (``paramagnetic'', shown in red).}
\label{fig:c_T_PD}
\end{center}
\end{figure}

In short, our results derive from the observation that in the adiabatic limit, i.e., taking the runtime $\tau$ of ARA to be large, the state of the system approaches the instantaneous equilibrium state $e^{-\beta H_S(t)}$ at the temperature $\beta^{-1} \equiv T$ of the environment.
We can then determine the performance of open-system ARA by calculating the \textit{finite}-temperature phase diagram of the model.
Analogous to closed-system annealing, the equilibrium state of $H_0$ can be prepared efficiently if there are paths in the phase diagram which avoid discontinuous transitions, although we must also check whether that equilibrium state is sufficiently close to the desired ground state (thus there are two failure modes for open-system ARA).
The result is shown in Fig.~\ref{fig:c_T_PD}, where we indicate whether ARA can locate the ground state as a function of temperature and initial guess $\{ \epsilon_j \}$ (specifically the fraction $c$ of spins that have the same orientation in the initial guess as in the ground state).
Note in particular that: i) if ARA succeeds at zero temperature, then it continues to succeed at low but non-zero temperature; ii) there is a small interval of $c$ in which ARA fails at zero temperature but succeeds for a range of non-zero temperatures.

The remainder of the paper is as follows.
We define the model and justify our claim about the equilibrium state in Sec.~\ref{sec:Thermodynamics}, then study the finite-temperature phase diagram in Sec.~\ref{sec:Dynamics}.
We summarize the conclusions in Sec.~\ref{sec:Conclusion}, and give a formal derivation of the mean-field dynamical equations in an appendix.

\section{Equilibration} \label{sec:Thermodynamics}

As do many theoretical studies of QA, we perform our analysis on the uniform $p$-spin model:
\begin{equation} \label{eq:p-spin model}
H_0 = -N \bigg( \frac{1}{N} \sum_{j=1}^N \hat{\sigma}_j^z \bigg)^p.
\end{equation}
Recall that the full ARA Hamiltonian $H_S(t)$ is given by Eq.~\eqref{eq:ARA Hamiltonian}, with parameters $s(t)$ and $\lambda(t)$ that depend on time $t \in [0, \tau]$ (with $\tau$ the runtime of the protocol).
Even though the ground state of $H_0$ is trivial (all $\hat{\sigma}_j^z = 1$), this model is a valuable test case for QA since (in the conventional case) it exhibits a discontinuous phase transition at a critical value of $s$~\cite{jorg2010energy,Bapst2012On}.
We fix $p = 3$ throughout this work for concreteness, but we expect the conclusions to hold for all odd $p$ (note that the ground state is unique for odd $p$ but doubly degenerate for even $p$).

The natural figure of merit for how close the system is to the ground state of $H_0$ is the $z$-magnetization $m(t) \equiv N^{-1} \sum_j \textrm{Tr}[ \rho(t) \hat{\sigma}_j^z ]$, where $\rho(t)$ is the density matrix of the system at time $t$.
Furthermore, due to the permutation symmetry among spins, all choices of $\{ \epsilon_j \}$ which have the same fraction of ones give equivalent $m(t)$ under ARA.
Denote that fraction by $c \equiv N^{-1} \sum_j \delta_{\epsilon_j, 1}$.
Thus the initial state has $m(0) = 2c - 1$, and ideally we would find that $m(\tau) \sim 1$ for $\tau$ scaling no larger than polynomial in $N$.
In the closed-system setting, by determining the ground-state phase diagram as a function of $(s, \lambda)$ and looking for paths which avoid phase transitions, it has been established that this is indeed the case~\cite{reverse_annealing_p_spin_model}.
However, it requires that $c$ be greater than a critical value --- i.e., that $\{ \epsilon_j \}$ be sufficiently close to the true ground state --- and even then, the successful paths $(s(t), \lambda(t))$ are non-trivial.

Following Ref.~\cite{standard_QA_outperform_ARA}, we incorporate environmental effects systematically by considering the system coupled to a bosonic bath.
The total Hamiltonian has the form
\begin{equation} \label{eq:system-bath hamiltonian}
H(t) = H_S(t) + H_B + H_{SB}
\end{equation}
where $H_S(t)$ is the full ARA Hamiltonian, $H_B$ is the Hamiltonian of the bath, and $H_{SB}$ gives the system-bath interaction.
Generically, the latter can be written
\begin{equation} \label{eq:interaction hamiltonian}
H_{SB} = g \sum_{\alpha} \hat{A}_\alpha \otimes \hat{B}_\alpha,
\end{equation}
where $\alpha$ is an abstract index labeling terms, $\hat{A}_\alpha$ and $\hat{B}_\alpha$ are dimensionless Hermitian operators acting on the system and bath respectively, and $g$ is an overall system-bath coupling strength.

In the weak-coupling (small-$g$) and adiabatic (large-$\tau$) limit, following a known derivation~\cite{quantum_markovian_master_equation}, the bath can be integrated out and the dynamics of the system is given by the master equation
\begin{equation} \label{eq:total Linbladian equation}
\partial_t \rho(t) = -i \big[ H_S(t) + H_{\text{LS}}(t), \rho(t) \big] + \mathcal{D} \big[ \rho(t) \big],
\end{equation}
where $H_{\text{LS}}(t)$ is the Lamb-shift operator:
\begin{equation} \label{eq:Lambshift}
H_{\text{LS}}(t) \equiv \sum_{\alpha \alpha'} \sum_{\omega} S_{\alpha \alpha'}(\omega) L_{\alpha, \omega}(t)^{\dagger} L_{\alpha', \omega}(t),
\end{equation}
and $\mathcal{D}$ is the dissipator superoperator:
\begin{equation} \label{eq:Dissipation}
\begin{aligned}
\mathcal{D} \big[ \rho(t) \big] \equiv \sum_{\alpha \alpha'} \sum_{\omega} &\gamma_{\alpha \alpha'}(\omega) \Big( L_{\alpha', \omega}(t) \rho(t) L_{\alpha, \omega}(t)^{\dagger} \\
&\quad - \frac{1}{2} \big\{ L_{\alpha, \omega}(t)^{\dagger} L_{\alpha',\omega}(t), \rho(t) \big\} \Big).
\end{aligned}
\end{equation}
In Eqs.~\eqref{eq:Lambshift} and~\eqref{eq:Dissipation}, $\omega$ runs over all differences between eigenvalues of $H_S(t)$, and the jump operator $L_{\alpha, \omega}(t)$ is given by~\cite{jump_operator_form_note}
\begin{equation} \label{eq:jump operators}
L_{\alpha, \omega}(t) = \sum_{a,b} \delta_{\varepsilon_b(t) - \varepsilon_a(t), \omega} \big| \varepsilon_a(t) \big> \big< \varepsilon_a(t) \big| \hat{A}_\alpha \big| \varepsilon_b(t) \big> \big< \varepsilon_b(t) \big|,
\end{equation}
where the sums run over all eigenstates of $H_S(t)$ and $\varepsilon_a(t), \varepsilon_b(t)$ denote the corresponding eigenvalues.
Lastly, the constants $\gamma_{\alpha \alpha'}(\omega)$ are the relaxation rates and $S_{\alpha \alpha'}(\omega)$ are their corresponding Hilbert transforms.
Since these are the only quantities deriving from the bath which enter into the master equation, we specify them directly rather than start with formulas for $H_B$ and $H_{SB}$.

Note that the jump operators depend on time only through $s(t)$ and $\lambda(t)$.
It thus stands to reason that in the adiabatic limit, where $s(t)$ and $\lambda(t)$ vary extremely slowly, the instantaneous steady state of the master equation --- i.e., the steady state holding $s$ and $\lambda$ fixed at their current values --- should be the equilibrium state $\rho \propto e^{-\beta H_S}$, where $\beta^{-1}$ is the temperature of the bath.
Indeed, one can confirm that the right-hand side of Eq.~\eqref{eq:total Linbladian equation} vanishes upon setting $\rho(t) \propto e^{-\beta H_S(t)}$, using only the property of detailed balance ($\gamma_{\alpha \alpha'}(\omega) = e^{\beta \omega} \gamma_{\alpha' \alpha}(-\omega)$).
Taking the argument one step further, we expect that as $s(t)$ and $\lambda(t)$ change over time (still in the adiabatic limit), the system will simply ``follow'' the instantaneous steady state, i.e., $\rho(t) \propto e^{-\beta H_S(t)}$ is in fact the solution to the master equation in the large-$\tau$ limit.
There is one essential caveat which we discuss in Sec.~\ref{sec:Dynamics}, but in the remainder of this section, we explicitly demonstrate this result for the $p$-spin model.

First, we finish specifying the properties of the bath.
We use an ``independent dephasing'' model in which each spin is coupled to an equivalent but independent bosonic bath through the operator $\hat{A}_{\alpha} = \hat{\sigma}_j^z$ (thus the index $\alpha$ simply runs over $j$).
The relaxation rates decouple and are equal ($\gamma_{jj'}(\omega) = \gamma(\omega) \delta_{jj'}$), and we use the form appropriate for an Ohmic bath:
\begin{equation} \label{eq:relaxation rate}
\gamma(\omega) = 2 \pi \eta \frac{\omega e^{-|\omega|/\omega_c}}{1 - e^{-\beta \omega}}.
\end{equation}
with $\eta$ the dimensionless coupling constant and $\omega_c$ the high-frequency cutoff.
Following Ref.~\cite{standard_QA_outperform_ARA}, we set $\eta = 10^{-3}$ and $\omega_c = 8 \pi$ throughout, but we do consider varying the bath temperature $\beta^{-1}$ (and indeed find that its value is extremely important).

Under this independent dephasing model, we can use a ``mean-field decoupling'' to determine the exact dynamics of the system quasi-analytically (in the thermodynamic limit).
A formal derivation is given in the appendix, but the result is straightforward to state: instead of evolving the many-body density matrix $\rho(t)$ under the interacting Hamiltonian $H_0$ (and all other terms), we can evolve separate $2 \times 2$ density matrices $\rho_j(t)$ for each spin under a self-consistent ``mean field'' $h(t)$ (and all other terms, including the interactions with the baths --- for this reason, it is important that we consider an independent bath for each spin).
More precisely, spin $j$ evolves under Hamiltonian
\begin{equation} \label{eq:effective H_j}
\begin{aligned}
\overline{H}_j(t) &= -\Big[ h(t)+ \big( 1 - s(t) \big) \big( 1 - \lambda(t) \big) \epsilon_j \Big]  \hat{\sigma}_j^z \\
&\qquad \qquad \qquad - \big( 1 - s(t) \big) \lambda(t) \hat{\sigma}_j^x,
\end{aligned}
\end{equation}
with the additional field given by
\begin{equation} \label{eq:h(t)}
h(t) = s(t) p m(t)^{p-1}.
\end{equation}
In addition to being the average magnetization in the original model, $m(t)$ must self-consistently also be the average magnetization for the uncoupled spins:
\begin{equation} \label{eq:z magnetization}
m(t) = \frac{1}{N} \sum_{j = 1}^N \text{Tr} \big[ \rho_j(t) \hat{\sigma}_j^z \big]. 
\end{equation}

Since spin $j$ still experiences the interaction with its bath~\cite{decoupling_note}, the evolution of $\rho_j(t)$ is given by a master equation analogous to Eq.~\eqref{eq:total Linbladian equation}:
\begin{equation} \label{eq:independent Lindblad equation}
\partial_t \rho_j(t) = -i \big[ \overline{H}_j(t) + {H}_{\textrm{LS},j}(t), \rho_j(t)\big] + \mathcal{D}_j \big[ \rho_j(t) \big].
\end{equation}
The Lamb-shift Hamiltonian and dissipator have the same form as in Eqs.~\eqref{eq:Lambshift} and~\eqref{eq:Dissipation}, except that now the eigenstates being summed over are those of the single-spin Hamiltonian $\overline{H}_j(t)$.
As $\overline{H}_j(t)$ is a $2 \times 2$ matrix, we can obtain explicit expressions for the allowed frequencies and jump operators --- see the appendix.

Two features of Eqs.~\eqref{eq:effective H_j} through~\eqref{eq:independent Lindblad equation} make them particularly easy to simulate.
First, $\overline{H}_j(t)$ differs between spins only in the value of $\epsilon_j$, the only possibilities for which are $1$ and $-1$.
Thus we do not need to evolve $N$ separate density matrices $\rho_j(t)$, but only two: the matrix $\rho_u(t)$ which gives the state for every spin having $\epsilon_j = 1$, and the matrix $\rho_d(t)$ which gives the state for every spin having $\epsilon_j = -1$.
Second, the self-consistency condition in Eq.~\eqref{eq:z magnetization} is straightforward to enforce due to causality: since $\rho_j(t)$ depends only on $h(t')$ for $t' < t$, we can determine $m(t)$ iteratively by, once it has been determined up to time $t$, fixing $h(t)$ via Eq.~\eqref{eq:h(t)}, evolving for a short timestep $\Delta t$ using that value for $h(t)$, then computing $m(t + \Delta t)$ via Eq.~\eqref{eq:z magnetization} (and then repeating).
We set $\Delta t = 0.1$ throughout.
A more precise statement of the algorithm is given at the end of the appendix.

There is one subtlety in interpreting the results of these calculations: the mean-field decoupling is valid only in the $N \rightarrow \infty$ limit, so we cannot determine finite-$N$ quantities such as the overlap between $\rho(\tau)$ and the ground state --- we can only compute the large-$N$ averages of quantities such as the magnetization $m(t)$.
That said, determining $m(t)$ is sufficient to assess how ARA will perform in approximate optimization --- whether it can produce states whose final energies are within a finite-percentage error of the ground state --- since the energy density of the $p$-spin model is a monotonic function of $m$: $\langle H_0 \rangle/N = -m^p$.

\begin{figure}[t]
\begin{center}
\includegraphics[width= 1.\linewidth, trim = 0 0 10 10, clip]{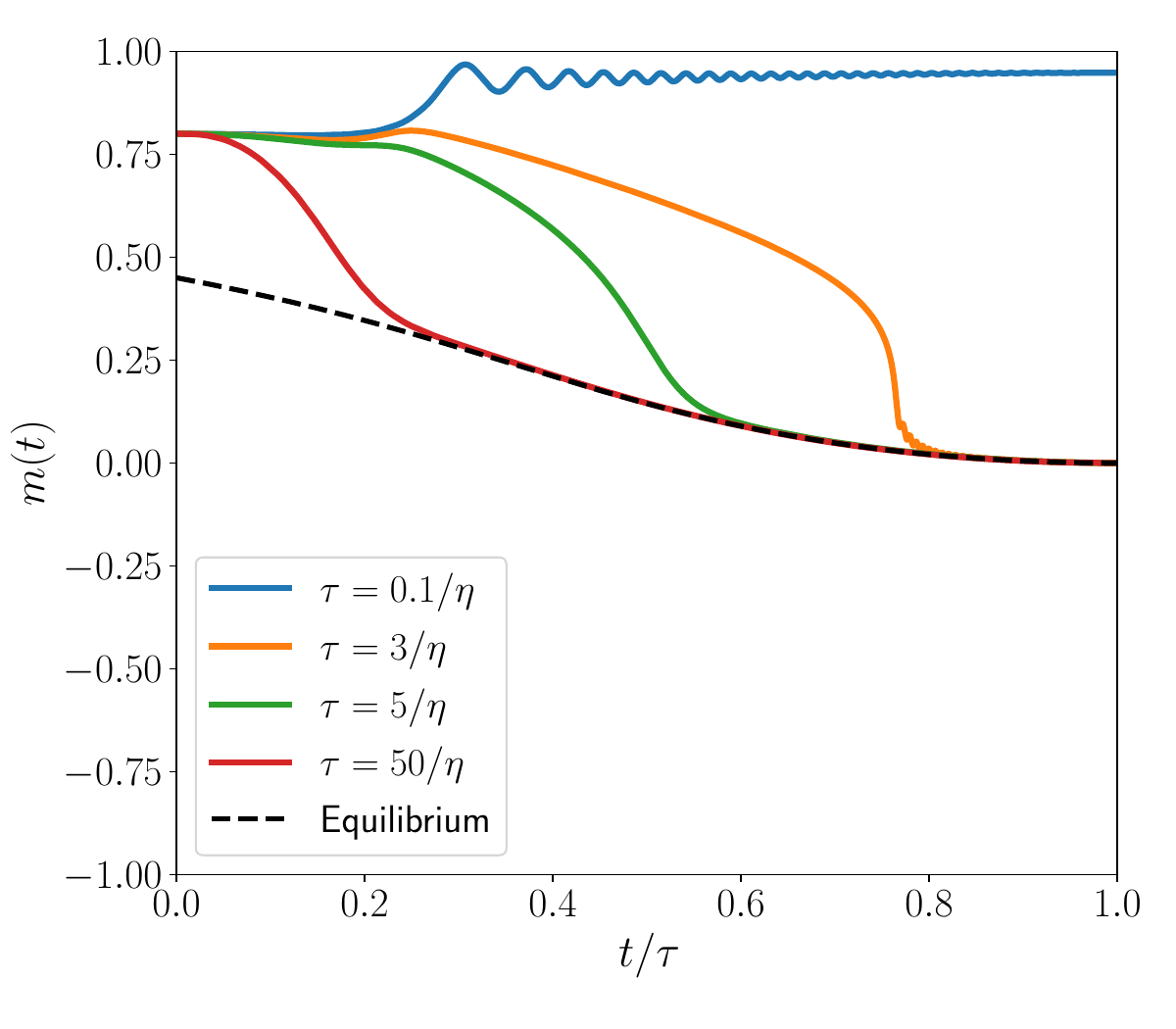}
\caption{The magnetization $m(t)$ during the ARA protocol for different runtimes $\tau$ (along the path $s(t) = \lambda(t) = t/\tau$). The black dashed line indicates the corresponding equilibrium value of the magnetization at that point in the protocol, i.e., the average using the state $e^{-\beta H_S(t)}$, with $\beta^{-1} = 1.57$. The initial state has $c = 0.9$.}
\label{fig:equilibriation time}
\end{center}
\end{figure}

Having fully described the setup, we can now demonstrate that the system does indeed follow the instantaneous equilibrium state $e^{-\beta H_S(t)}$ when $s(t)$ and $\lambda(t)$ are varied sufficiently slowly.
Fig.~\ref{fig:equilibriation time} gives a representative example.
The bath temperature $T = \beta^{-1}$ is set to $1.57$ (the same value used in Ref.~\cite{standard_QA_outperform_ARA}), the protocol follows the path $s(t) = \lambda(t) = t/\tau$, and trajectories $m(t)$ are shown for four choices of $\tau$.
Since the dissipator comes with an overall factor of $\eta$, it is natural to measure $\tau$ in units of $1/\eta$.
For $\tau = 0.1/\eta$, which is a sufficiently short protocol that the bath has only a weak effect, we see oscillations in $m(t)$ characteristic of (approximately) closed-system dynamics, and the trajectory ends at a value reasonably close to $1$.
The situation is very different for $\tau = 3/\eta$: the magnetization instead relaxes to the equilibrium value (whose determination is discussed below).
As $\tau$ increases further, the magnetization reaches the equilibrium value earlier and earlier in the protocol, and once it does, it indeed tracks the equilibrium value from that point on.

Lastly, note that these results confirm the conclusion of Ref.~\cite{standard_QA_outperform_ARA} via complementary means: for this choice of temperature, since the equilibrium magnetization (and thus $m(t)$ in the adiabatic limit) tends to zero at the end of the protocol, ARA fails to locate the desired ground state and shows no advantage over conventional QA.

\section{Phase diagrams} \label{sec:Dynamics}

Since the system follows the instantaneous equilibrium state in the adiabatic limit, we can predict the performance of ARA by determining the \textit{finite}-temperature phase diagram, specifically at the temperature of the bath.
One consequence of this is clear: ARA will fail if the equilibrium state at the end of the protocol is in the paramagnetic phase ($m = 0$).
Note that the final state is nothing more than the Boltzmann distribution for the original problem: $\rho(\tau) \propto e^{-\beta H_0}$.
More generally, ARA fails if the configurations obtained by sampling that state do not lie in the ``basin of attraction'' of the ground state, i.e., if simple classical algorithms such as steepest descent cannot locate the ground state starting from the ARA output.
For the $p$-spin model considered here, this amounts simply to whether $e^{-\beta H_0}$ is in the paramagnetic or ferromagnetic phase.

\begin{figure}[t]
\begin{center}
\includegraphics[width= 1.\linewidth, trim = 0 0 10 10, clip]{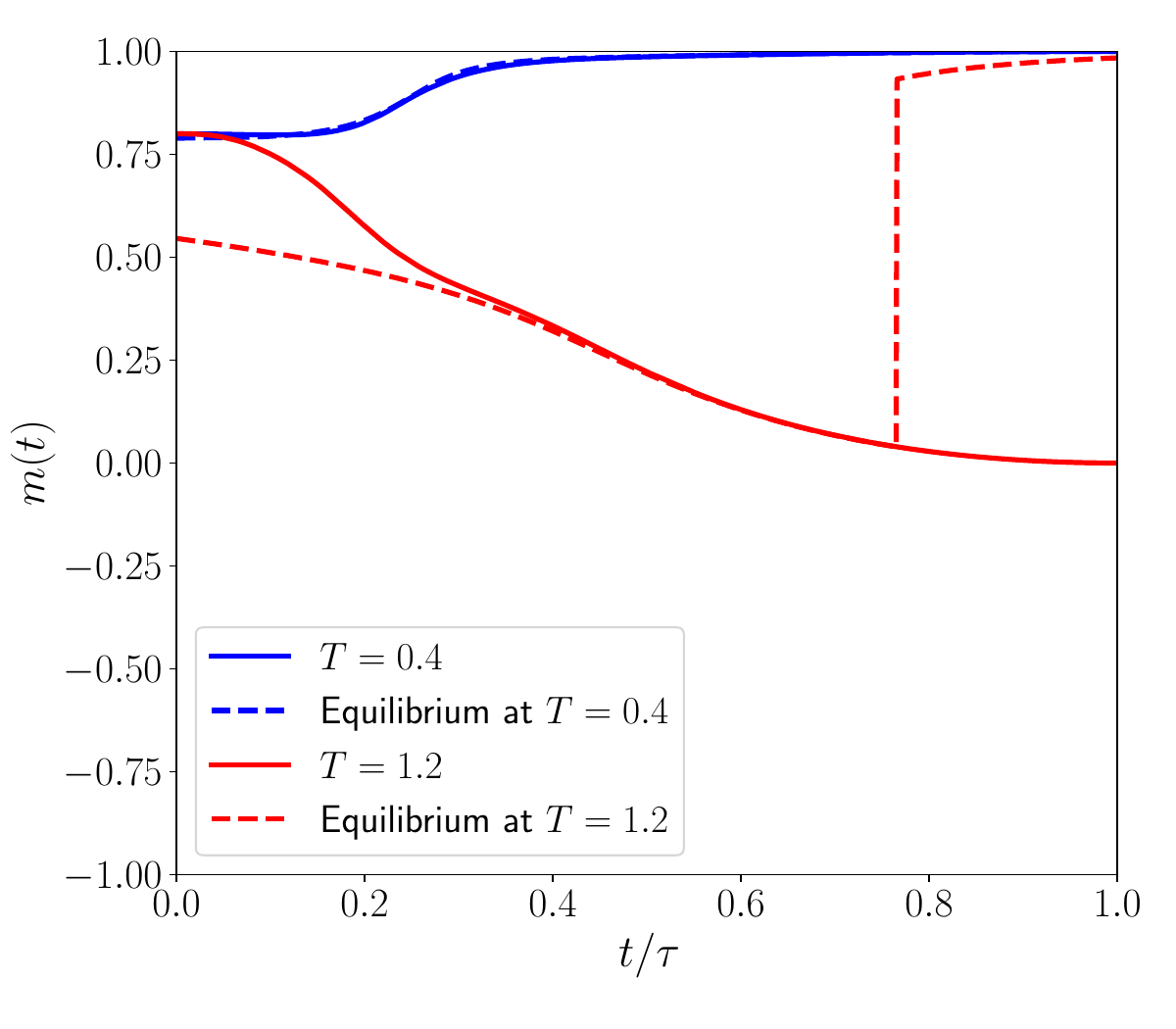}
\caption{The magnetization $m(t)$ during the ARA protocol at two different temperatures $T = \beta^{-1}$. Solid lines show the true dynamical evolution of $m(t)$, while the dashed lines indicate the corresponding equilibrium values at that point in the protocol, i.e., the average using the state $e^{-\beta H_S(t)}$ for the two temperatures in question. Runtime is $\tau = 50/\eta$, the path is $s(t) = \lambda(t) = t/\tau$, and $c = 0.9$. The discontinuity in the red dashed line indicates that the protocol crosses a discontinuous phase transition at that temperature.}
\label{fig:thermal and dynamical comparison at 2 temps}
\end{center}
\end{figure}

Beyond this, ARA also requires that one follow a path in the phase diagram which avoids transitions, exactly as closed-system ARA does.
This is because there is a subtlety to our earlier claim: if the protocol crosses a discontinuous transition, where the equilibrium state jumps between two local minima of the free energy landscape, then the system will only follow the equilibrium state on a timescale exponential in $N$ (at least in infinite-range models such as ours, where the free energy barrier between minima is extensive).
On all shorter timescales, the system will follow whichever local minimum it begins in, even as that becomes a metastable state rather than the true equilibrium state.

Fig.~\ref{fig:thermal and dynamical comparison at 2 temps} demonstrates this by comparing the trajectories $m(t)$ at two temperatures.
Both follow the same protocol ($s(t) = \lambda(t) = t/\tau$), but no transition is crossed at the lower temperature whereas one is at the higher temperature (the corresponding phase diagrams are shown as the orange and red lines respectively in Fig.~\ref{fig:Finite-temperature phase diagram} below).
In the former case, $m(t)$ simply follows the equilibrium value throughout the protocol.
In the latter case, it does so only until the transition is reached --- at that point, there is a jump in the equilibrium value, but $m(t)$ instead follows a smooth continuation of its previous trajectory~\cite{order_limits_note}.

In short, the system can only follow the equilibrium state if the ARA protocol does not cross any discontinuous phase transitions.
Thus to determine whether ARA succeeds, we calculate the finite-temperature phase diagram and look for two features:
\begin{itemize}
\item The system is in the ferromagnetic phase at $s = 1$.
\item There are paths from $s = 0$ to $s = 1$ which avoid transitions.
\end{itemize}
If both of these conditions are satisfied, then ARA is successful.

Most of the work to determine the ARA phase diagram for the $p$-spin model has already been done in Ref.~\cite{reverse_annealing_p_spin_model}.
It is derived there that the equilibrium value of $m$ is given by the solution to the equation
\begin{widetext}
\begin{equation} \label{eq:self_consistent_magnetization}
m = \frac{1}{N} \sum_{j=1}^N \frac{spm^{p-1} + (1 - s) (1 - \lambda) \epsilon_j}{\sqrt{\big[ spm^{p-1} + (1 - s) (1 - \lambda) \epsilon_j \big]^2 + (1 - s)^2 \lambda^2}} \tanh{\beta \sqrt{\big[ spm^{p-1} + (1 - s) (1 - \lambda) \epsilon_j \big]^2 + (1 - s)^2 \lambda^2}},
\end{equation}
and the resulting free energy is
\begin{equation} \label{eq:general_p_spin_free_energy}
f = s(p-1)m^p - \frac{1}{N \beta} \sum_{j=1}^N \log{2 \cosh{\beta \sqrt{\big[ spm^{p-1} + (1 - s) (1 - \lambda) \epsilon_j \big]^2 + (1 - s)^2 \lambda^2}}}.
\end{equation}
\end{widetext}
If there are multiple solutions to Eq.~\eqref{eq:self_consistent_magnetization}, that which gives the lowest value for Eq.~\eqref{eq:general_p_spin_free_energy} is chosen.
Whereas Ref.~\cite{reverse_annealing_p_spin_model} then takes the $T \rightarrow 0$ ($\beta \rightarrow \infty$) limit of these expressions, here we keep $T$ as a parameter.

\begin{figure}[t]
\begin{center}
\includegraphics[width= 1\linewidth, trim = 20 40 0 30, clip]{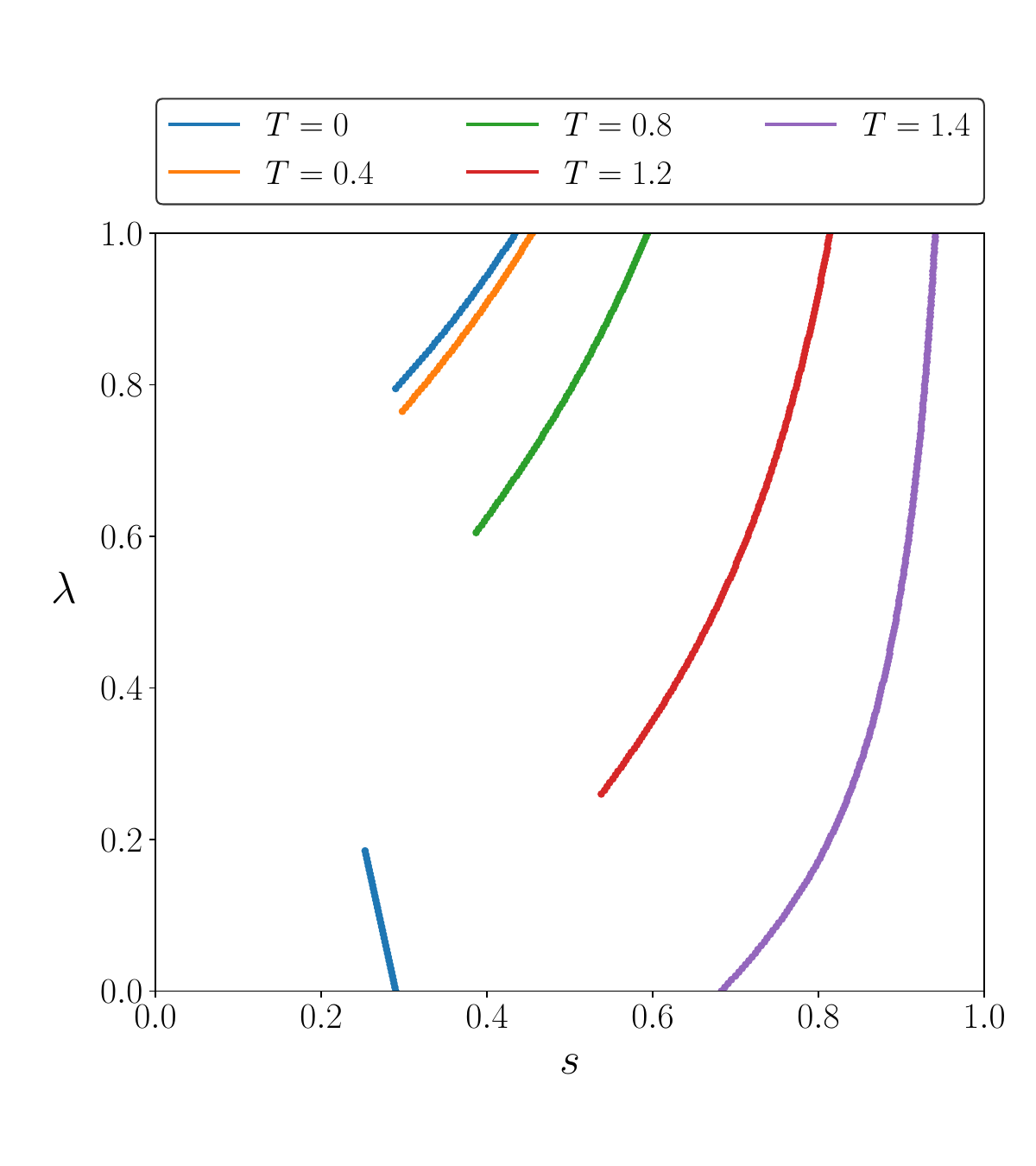}
\caption{Equilibrium phase boundaries in the $s$-$\lambda$ plane for the ARA Hamiltonian $H_S(s, \lambda)$ (Eq.~\eqref{eq:ARA Hamiltonian}), at various temperatures $T$ (with $c = 0.9$). All lines indicate discontinuous phase transitions at which there is a jump in the magnetization (each boundary is identified by where the magnetization jumps by more than $0.05$ from one value of $s$ to the next).}
\label{fig:Finite-temperature phase diagram}
\end{center}
\end{figure}

Fig.~\ref{fig:Finite-temperature phase diagram} shows the phase boundaries of the $p$-spin model for a range of temperatures (setting $c = 0.9$).
At $T = 0$, we confirm the result of Ref.~\cite{reverse_annealing_p_spin_model} that there are two separate phase boundaries and an opening between them --- there are paths from $s = 0$ to $s = 1$ which avoid transitions, and ARA is successful.
As the temperature increases, the lower phase boundary quickly disappears but the upper boundary extends downwards.
It has reached the bottom of the phase diagram by $T = 1.4$, and there are no longer any transition-avoiding paths --- ARA fails at and above this temperature (more precise calculations give that ARA fails for $T > T_{c1} \approx 1.37$).

We check the equilibrium phase at the end of the protocol by setting $s = 1$ in the above equations --- they reduce to
\begin{equation} \label{eq:classical_self_consistent_magnetization}
m = \tanh{\beta p m^{p-1}},
\end{equation}
\begin{equation} \label{eq:classical_general_p_spin_free_energy}
f = (p-1) m^p - \beta^{-1} \log{2 \cosh{\beta pm^{p-1}}}.
\end{equation}
These show the textbook behavior that one expects for discontinuous transitions in mean-field models (recall that we set $p = 3$): Eq.~\eqref{eq:classical_self_consistent_magnetization} has only the single solution $m = 0$ for temperatures above a critical value, but two additional (positive) solutions below that value, and the largest solution is the correct one when it exists.
The critical temperature is determined numerically to be $T_{c2} \approx 1.49$ --- the system is paramagnetic for $T > T_{c2}$ and ferromagnetic for $T < T_{c2}$.

\begin{figure*}[t]
\begin{center}
\includegraphics[width= 1.01\linewidth, trim = 27 0 0 0, clip]{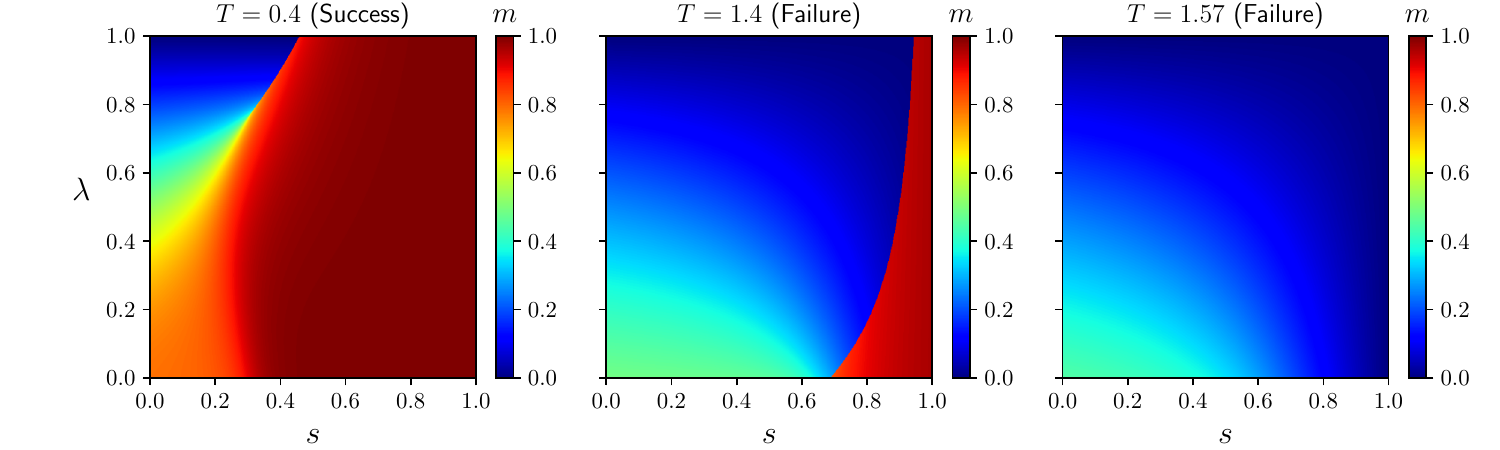}
\caption{Equilibrium phase diagram for the ARA Hamiltonian $H_S(s, \lambda)$ (Eq.~\eqref{eq:ARA Hamiltonian}), with the color indicating the equilibrium value of the magnetization, at three representative temperatures (with $c = 0.9$). At $T = 0.4$ (left), ARA is successful since there are paths from $(0, 0)$ to $(1, 1)$ that do not cross discontinuous phase transitions, as evidenced by the color changing smoothly. At $T = 1.4$ (middle), ARA fails since there are no longer any such paths. At $T = 1.57$ (right), ARA fails since the final state is in the paramagnetic phase (average magnetization of zero).}
\label{fig:phase diagrams different temps}
\end{center}
\end{figure*}

It is interesting to note that the temperature at which transition-avoiding paths disappear is \textit{not} the same temperature at which the final state becomes paramagnetic.
We specifically find that $T_{c1} < T_{c2}$.
Nonetheless, it is important to keep both failure mechanisms in mind, for we observe that no phase transitions at all occur when $T > T_{c2}$.
If one were focusing solely on whether transition-avoiding paths exist, and not checking the magnetization of the final state, one might erroneously believe ARA to be successful at $T > T_{c2}$.
We summarize the three possible types of phase diagrams in Fig.~\ref{fig:phase diagrams different temps}.
In only the first case, at the lowest temperatures, does ARA succeed.

Lastly, we determine for which values of $(c, T)$ does ARA succeed, and by which of the two mechanisms does it fail if not~\cite{phase_diagram_note}.
The results are shown above in Fig.~\ref{fig:c_T_PD}.
The most striking feature is that, remarkably, there is an intermediate range of $c$ (between roughly $0.71$ and $0.74$) where the bath \textit{benefits} ARA: the protocol fails to reach the ground state at $T = 0$ but succeeds at intermediate temperatures.

While further investigation is warranted, particularly of whether this phenomenon may hold in genuinely hard spin-glass problems, we can provide a partial explanation using the interpretation of ARA in Ref.~\cite{Baldwin2025Simulated}.
In closed-system ARA, the transition when increasing $s$ at small $\lambda$ corresponds to the system jumping from (approximately) the initial state to the $p$-spin ground state, while the transition at large $\lambda$ is between a paramagnetic state and the $p$-spin ground state.
The main effect of the transverse field is to provide fluctuations that can kick the system over barriers in the energy landscape (a positive effect) but also drive the system towards a paramagnetic state (a negative effect)~\cite{Baldwin2025Simulated}.
Note that the bath provides additional fluctuations as quantified by the temperature.
Thus increasing $T$ should suppress the small-$\lambda$ transition but extend the large-$\lambda$ transition, which is what we observe in Fig.~\ref{fig:Finite-temperature phase diagram}.
We also observe that the small-$\lambda$ transition seems to be more sensitive to temperature --- it disappears quite rapidly as $T$ increases, whereas the large-$\lambda$ transition is slower to extend.
This suggests that even if there is no transition-avoiding path at $T = 0$, an opening may emerge when increasing $T$, and this is indeed what we see in Fig.~\ref{fig:c_T_PD}.

Beyond this, the other features of Fig.~\ref{fig:c_T_PD} are as expected.
At $c = 1/2$, where the initial state is uncorrelated with the $p$-spin ground state, ARA fails at all temperatures --- this failure persists for $c$ close to but slightly above $1/2$.
At the other end, if $c$ is sufficiently large that closed-system ARA succeeds, then open-system ARA will continue to succeed at low but non-zero temperature, and the breakdown temperature for ARA increases as $c$ nears $1$.
Also note that the temperature $T_{c2}$ above which the final state is paramagnetic is a property solely of the $p$-spin model, and thus is independent of $c$.
It sets an upper limit to the functioning of ARA regardless of the initial state.

\section{Conclusion} \label{sec:Conclusion}

In the closed-system setting, ARA has been shown to be capable of circumventing the discontinuous phase transitions that plague conventional QA, at least in solvable mean-field models~\cite{reverse_annealing_p_spin_model}.
Yet recent numerical results have suggested that ARA may lose its advantage in the presence of noise~\cite{standard_QA_outperform_ARA}.
Here we have shown (again in the solvable $p$-spin model) that the situation is more encouraging: open-system ARA can in fact retain its exponential speedup over conventional QA, although the temperature of the environment must be sufficiently low.
Quite interestingly, we find that it is even possible for a low but non-zero temperature to \textit{improve} ARA, rendering it efficient in cases where it would have been exponentially slow at zero temperature.

These results stem from the fact that in the adiabatic limit (runtime $\tau \rightarrow \infty$), the system relaxes into the instantaneous equilibrium state $e^{-\beta H_S(t)}$, with $\beta^{-1}$ the temperature of the bath.
Thus determining the behavior of ARA reduces to determining the finite-temperature phase diagram of the model.
Much as for closed-system ARA, the equilibrium state of $H_0$ can only be reached efficiently when following a path from $(s, \lambda) = (0, 0)$ to $(1, 1)$ that does not cross any discontinuous phase transitions.
We find that the (discontinuous) paramagnetic-ferromagnetic transition at large $\lambda$ extends towards smaller $\lambda$ as the temperature increases, and thus transition-avoiding paths will ultimately disappear at high $T$ even if they exist at low $T$.
On the other hand, the (also discontinuous) small-$\lambda$ transition between the initial guess and the true ground state quickly disappears with temperature --- this is what makes it possible for ARA to benefit from non-zero temperature.

Another failure mechanism for ARA is that the equilibrium state at the end of the protocol may simply be paramagnetic (or more generally, outside the ``basin of attraction'' of the desired ground state).
This sets an initial-state-independent upper limit to the temperatures at which ARA can succeed.
The temperature used in Ref.~\cite{standard_QA_outperform_ARA} in fact lies above this limit, which explains the discouraging results seen in that study.

In a sense, this work is a natural companion to the concurrent Ref.~\cite{Baldwin2025Simulated}: both study the relationship between quantum and thermal fluctuations in ARA.
Yet whereas Ref.~\cite{Baldwin2025Simulated} considers them one at a time, so as to motivate a classical ``simulated'' analogue of ARA, here we consider the joint effect of the two --- quantum-mechanical ARA in the presence of thermal fluctuations.
An interesting direction for future work would be treating the temperature not as a fixed parameter but as another knob to vary, much as Ref.~\cite{Baldwin2025Simulated} does, to see whether there may be paths in the $s$-$\lambda$-$T$ hyperplane that provide additional improvement over ARA.

Another important future direction is to go beyond the weak system-bath coupling considered here, especially since it has been argued that the weak-coupling master equation is not particularly accurate in describing present-day quantum annealers~\cite{Bando2022Breakdown}.
To a certain degree, the intuition used here still holds at larger coupling: the system will still come to equilibrium with the bath, the only difference being that the equilibrium state is no longer $e^{-\beta H_S(t)}$ but rather the partial trace of $e^{-\beta [H_S(t) + H_B + H_{SB}]}$.
Thus one must study the finite-temperature phase diagram of a (many-body) spin-boson model in order to predict the performance of ARA.
For the $p$-spin model, such a study is conceivable: the mean-field decoupling remains valid, and so the problem reduces to investigating the standard spin-boson model in a self-consistent field.
A host of numerical methods exist for the spin-boson problem and could be brought to bear, but we leave this for future work.

\section{Acknowledgements} \label{sec:acknowledgements}

It is a pleasure to thank Philip Crowley and Mohammad Maghrebi for valuable discussions and insights concerning open-system dynamics.
This work was supported by the U.S. National Science Foundation under award No.~2508604, as well as start-up funds from Michigan State University.

\bibliography{citation}

\begin{thebibliography}{58}
\bibitem [{jum()}]{jump_operator_form_note}%
  \BibitemOpen
  \href@noop {} {}\bibinfo {note} {This form for the jump operators only holds if both the eigenvalues and the transitions are non-degenerate: not only must $\varepsilon_a \neq \varepsilon_b$ for $a \neq b$, but also for any non-zero frequency $\omega$, there must be only one pair of eigenvalues $(\varepsilon_a, \varepsilon_b)$ with $\varepsilon_b - \varepsilon_a = \omega$. If this condition holds, and it does in our case after making the mean-field decoupling (which we do before deriving the master equation), then Eq.~\eqref{eq:jump operators} is a compact way to write the jump operators for both zero and non-zero frequency.}\BibitemShut {Stop}%
\bibitem [{dec()}]{decoupling_note}%
  \BibitemOpen
  \href@noop {} {}\bibinfo {note} {Conceptually, the order in which we carry out the analysis is important. We first perform the mean-field decoupling, while still considering the system-bath Hamiltonian as in Eq.~\eqref{eq:system-bath hamiltonian}. Since each spin interacts with a separate bath, this reduces the Hamiltonian to the form $H(t) = \sum_j [\overline{H}_j(t) + H_{B,j} + H_{SB,j}]$, with no remaining coupling between spins. We then follow the standard derivation of the master equation, separately for each spin, to arrive at Eq.~\eqref{eq:independent Lindblad equation}.}\BibitemShut {Stop}%
\bibitem [{ord()}]{order_limits_note}%
  \BibitemOpen
  \href@noop {} {}\bibinfo {note} {Since our mean-field calculations to derive the single-spin master equation involve taking $N \rightarrow \infty$, all values of $\tau$ that we use are necessarily $O(1)$ with respect to $N$. Thus there is no possibility of seeing the system thermally activate over an extensive free energy barrier.}\BibitemShut {Stop}%
\bibitem [{pha()}]{phase_diagram_note}%
  \BibitemOpen
  \href@noop {} {}\bibinfo {note} {As in Fig.~\ref{fig:Finite-temperature phase diagram}, we determine the phase boundaries for a given $(c, T)$ --- from which it is straightforward to check whether transition-avoiding paths exist --- by looking for points at which the magnetization jumps by more than $0.05$ from one value of $s$ to the next. This is admittedly crude, so we have confirmed that our main conclusion --- there exist values of $c$ for which ARA fails at zero temperature but succeeds at a non-zero temperature --- holds more systematically by plotting Eq.~\eqref{eq:self_consistent_magnetization} as we vary parameters. At $(c, T) = (0.72, 0)$, the solution has a discontinuity while increasing $s$ for all $\lambda$. At $(c, T) = (0.72, 0.6)$, the solution is smooth along $s = \lambda$.}\BibitemShut {Stop}%
\end{thebibliography}

\begin{widetext}
\appendix
\section{Dynamics for the noisy ferromagnetic $p$-spin model under ARA} \label{sec:dynamical mean-field derivation}

Here we derive the mean-field equations for the dynamics of the magnetization $m(t)$ under open-system ARA.
We start from the standard Hamiltonian for a system coupled to a bath:
\begin{equation} \label{eq:generic_system_bath_Hamiltonian_form}
H(t) = H_S(t) + H_B + H_{SB},
\end{equation}
where $H_S(t)$ is the ARA Hamiltonian for the $p$-spin model:
\begin{equation} \label{eq:p_spin_ARA_Hamiltonian}
H_S(t) = -N s(t) \bigg( \frac{1}{N} \sum_{j=1}^N \hat{\sigma}_j^z \bigg)^p - \big( 1 - s(t) \big) \big( 1 - \lambda(t) \big) \sum_{j=1}^N \epsilon_j \hat{\sigma}_j^z - \big( 1 - s(t) \big) \lambda(t) \sum_{j=1}^N \hat{\sigma}_j^x,
\end{equation}
and we take the bath to consist of separate sets of harmonic oscillators coupled to each spin via the $\hat{\sigma}_j^z$ operator:
\begin{equation} \label{eq:bosonic_bath_definition}
H_B = \sum_{j=1}^N \sum_k \omega_k \hat{b}_{jk}^{\dagger} \hat{b}_{jk}, \qquad H_{SB} = g \sum_{j=1}^N \sum_k \kappa_k \hat{\sigma}_j^z \big( \hat{b}_{jk} + \hat{b}_{jk}^{\dagger} \big),
\end{equation}
where $\omega_k$ is the frequency of the $k$'th oscillator and $\kappa_k$ is its coupling coefficient ($g$ is an overall coupling strength).
Note that even though $\omega_k$ and $\kappa_k$ are independent of $j$, we still have separate oscillators for each spin.
As discussed in the main text, the derivation of a master equation for the system alone yields a Lindbladian that depends on the bath frequencies and couplings only through a single function $\gamma(\omega)$ (given below), so we specify $\gamma(\omega)$ directly rather than assign values to $\omega_k$ and $\kappa_k$.

The focus of this appendix is the ``mean-field decoupling'' that replaces the interactions between spins in Eq.~\eqref{eq:p_spin_ARA_Hamiltonian} by a self-consistent ``mean field'' $h(t)$.
Each spin then evolves independently under its own Hamiltonian $\overline{H}_j(t)$ and coupled to its own bath of oscillators, so we can use the known derivation of the master equation~\cite{quantum_markovian_master_equation} (for a \textit{single} spin coupled to a bath) to obtain the Lindbladian given in the main text.

Let $\rho(t)$ denote the joint system-bath density matrix at time $t$.
As is standard, we take the initial density matrix to factor: $\rho(0) = \rho_S(0) \otimes \rho_B(0)$, with $\rho_S(0)$ being the initial guess $\{ \epsilon_j \}$ (so $\rho_S(0) = \bigotimes_j | \epsilon_j \rangle \langle \epsilon_j |$) and $\rho_B(0)$ being the thermal state at temperature $\beta^{-1}$ (so $\rho_B(0) \propto e^{-\beta H_B}$).
The density matrix does not factor at later times, but we are only interested in the reduced density matrix for the system, $\rho_S(t) \equiv \textrm{Tr}_B \rho(t)$, and specifically the average magnetization, $m(t) \equiv N^{-1} \sum_j \textrm{Tr}_S [\rho_S(t) \hat{\sigma}_j^z]$.
We determine $m(t)$ by considering the Keldysh generating functional:
\begin{equation} \label{eq:Keldysh_generating_functional_definition}
\mathcal{Z} = \textrm{Tr} \Big( \mathcal{T} e^{-i \int_0^{\tau} \textrm{d}t H(t)} \Big) \rho(0) \Big( \mathcal{T} e^{-i \int_0^{\tau} \textrm{d}t H(t)} \Big)^{\dagger},
\end{equation}
where $\mathcal{T}$ denotes time-ordering (and $\tau$ is the total runtime of the ARA protocol).
Since the argument of the trace is equivalently $\rho(\tau)$, clearly $\mathcal{Z} = 1$, so $\mathcal{Z}$ itself is of no interest.
Yet we construct a path-integral representation of $\mathcal{Z}$ nonetheless, since a byproduct of that construction is a means of determining $m(t)$, as we discuss below.

To help with notation, write the full system-bath Hamiltonian as $H(t) = A(t) + \sum_j B_j(t)$, where
\begin{equation} \label{eq:full_Hamiltonian_decomposition_A}
A(t) \equiv -N s(t) \bigg( \frac{1}{N} \sum_{j=1}^N \hat{\sigma}_j^z \bigg)^p,
\end{equation}
\begin{equation} \label{eq:full_Hamiltonian_decomposition_B}
B_j(t) \equiv -\big( 1 - s(t) \big) \big( 1 - \lambda(t) \big) \epsilon_j \hat{\sigma}_j^z - \big( 1 - s(t) \big) \lambda(t) \hat{\sigma}_j^x + \sum_k \omega_k \hat{b}_{jk}^{\dagger} \hat{b}_{jk} + g \sum_k \kappa_k \hat{\sigma}_j^z \big( \hat{b}_{jk} + \hat{b}_{jk}^{\dagger} \big).
\end{equation}
Then apply a standard Suzuki-Trotter decomposition to each time-ordered exponential in $\mathcal{Z}$:
\begin{equation} \label{eq:Suzuki_Trotter_decomposition_v1}
\begin{aligned}
\mathcal{T} e^{-i \int_0^{\tau} \textrm{d}t \big[ A(t) + \sum_j B_j(t) \big]} &\sim e^{-i \big[ A(\tau) + \sum_j B_j(\tau) \big] \Delta t} \cdots e^{-i \big[ A(\Delta t) + \sum_j B_j(\Delta t) \big] \Delta t} e^{-i \big[ A(0) + \sum_j B_j(0) \big] \Delta t} \\
&\sim e^{-i A(\tau) \Delta t} \bigg( \prod_j e^{-i B_j(\tau) \Delta t} \bigg) \cdots e^{-i A(\Delta t) \Delta t} \bigg( \prod_j e^{-i B_j(\Delta t) \Delta t} \bigg) e^{-i A(0) \Delta t} \bigg( \prod_j e^{-i B_j(0) \Delta t} \bigg),
\end{aligned}
\end{equation}
where $\Delta t$ is a small (ideally infinitesimal) timestep.
Insert resolutions of the identity between all exponentials:
\begin{equation} \label{eq:identity_resolution}
\sum_{\{ \sigma_j \}, \{ \psi_j \}} \big| \{ \sigma_j(t) \}, \{ \psi_j(t) \} \big> \big< \{ \sigma_j(t) \}, \{ \psi_j(t) \} \big| = 1,
\end{equation}
where we sum over joint basis states for both spins and oscillators --- we choose the $z$-basis $\{ \sigma_j(t) \} \in \{1, -1\}^N$ for the spins, and any product basis $\{ \psi_j(t) \}$ for the oscillators will do (it need not even be a tensor product with respect to $k$).
Here $t$ is an additional index for the basis states --- we insert Eq.~\eqref{eq:identity_resolution} between the factors at time $t$ and $t + \Delta t$, doing so separately for each $t$.
Note that $A(t)$ is diagonal in the $z$-basis and acts as the identity on the oscillators:
\begin{equation} \label{eq:p_spin_basis_action}
e^{-i A(t) \Delta t} \big| \{ \sigma_j(t) \}, \{ \psi_j(t) \} \big> = \exp{\bigg[ iN s(t) \bigg( \frac{1}{N} \sum_{j=1}^N \sigma_j(t) \bigg)^p \Delta t \bigg]} \big| \{ \sigma_j(t) \}, \{ \psi_j(t) \} \big>,
\end{equation}
while the matrix elements of $\prod_j e^{-i B_j(t) \Delta t}$ factor across $j$.
Thus the time-ordered exponential becomes
\begin{equation} \label{eq:Suzuki_Trotter_decomposition_v2}
\begin{aligned}
\mathcal{T} e^{-i \int_0^{\tau} \textrm{d}t \big[ A(t) + \sum_j B_j(t) \big]} &\sim \sum_{\{ \sigma_j(t) \}, \{ \psi_j(t) \}} \exp{\bigg[ iN \sum_t s(t) \bigg( \frac{1}{N} \sum_{j=1}^N \sigma_j(t) \bigg)^p \Delta t \bigg]} \\
&\qquad \cdot \prod_j \big| \sigma_j(\tau), \psi_j(\tau) \big> \bigg( \prod_t \big< \sigma_j(t + \Delta t), \psi_j(t + \Delta t) \big| e^{-i B_j(t) \Delta t} \big| \sigma_j(t), \psi_j(t) \big> \bigg) \big< \sigma_j(0), \psi_j(0) \big|,
\end{aligned}
\end{equation}
and $\mathcal{Z}$, which incorporates two copies of Eq.~\eqref{eq:Suzuki_Trotter_decomposition_v2}, becomes
\begin{equation} \label{eq:Keldysh_generating_function_Suzuki_Trotter}
\begin{aligned}
\mathcal{Z} &\sim \sum_{\{ \sigma_j^{\pm}(t) \}, \{ \psi_j^{\pm}(t) \}} \exp{\bigg[ iN \sum_t s(t) \bigg( \frac{1}{N} \sum_{j=1}^N \sigma_j^+(t) \bigg)^p \Delta t - iN \sum_t s(t) \bigg( \frac{1}{N} \sum_{j=1}^N \sigma_j^-(t) \bigg)^p \Delta t  \bigg]} \\
&\qquad \qquad \cdot \bigg( \prod_{j,t} \big< \sigma_j^+(t + \Delta t), \psi_j^+(t + \Delta t) \big| e^{-i B_j(t) \Delta t} \big| \sigma_j^+(t), \psi_j^+(t) \big> \big< \sigma_j^-(t), \psi_j^-(t) \big| e^{i B_j(t) \Delta t} \big| \sigma_j^-(t + \Delta t), \psi_j^-(t + \Delta t) \big> \bigg) \\
&\qquad \qquad \qquad \qquad \cdot \bigg( \prod_j \big< \sigma_j^+(0) \big| \epsilon_j \big> \big< \epsilon_j \big| \sigma_j^-(0) \big> \big< \psi_j^+(0) \big| e^{-\beta \sum_k \omega_k \hat{b}_{jk}^{\dagger} \hat{b}_{jk}} \big| \psi_j^-(0) \big> \delta_{\sigma_j^+(\tau), \sigma_j^-(\tau)} \delta_{\psi_j^+(\tau), \psi_j^-(\tau)} \bigg).
\end{aligned}
\end{equation}
Note the additional index $\pm$ for the basis states, indicating whether they come from the left exponential ($+$) or the right exponential ($-$) in Eq.~\eqref{eq:Keldysh_generating_functional_definition}.
We have also used the specific form of the initial density matrix $\rho(0)$.
The lower lines of Eq.~\eqref{eq:Keldysh_generating_function_Suzuki_Trotter} factor across $j$, so denote the product of all terms involving spin $j$ and bath $j$ by $\mathcal{B}_j$ (i.e., write the lower lines simply as $\prod_j \mathcal{B}_j$).
In what follows, the explicit expression for $\mathcal{B}_j$ is irrelevant --- we need only remember that $\mathcal{B}_j$ is the collection of terms resulting from the Suzuki-Trotter decomposition of evolution under $B_j(t)$.

To proceed, introduce $\delta$-functions setting $\sum_j \sigma_j^r(t) = Nm^r(t)$ (where $r \in \{+, -\}$):
\begin{equation} \label{eq:Keldysh_top_line_simplification_v1}
\begin{aligned}
\mathcal{Z} &= \sum_{\{ \sigma_j^{\pm}(t) \}, \{ \psi_j^{\pm}(t) \}} \exp{\bigg[ iN \sum_t s(t) \bigg( \frac{1}{N} \sum_{j=1}^N \sigma_j^+(t) \bigg)^p \Delta t - iN \sum_t s(t) \bigg( \frac{1}{N} \sum_{j=1}^N \sigma_j^-(t) \bigg)^p \Delta t \bigg]} \prod_j \mathcal{B}_j \\
&= \int \prod_{t,r} \textrm{d}m^r(t) \exp{\bigg[ iN \sum_t \Big( s(t) m^+(t)^p - s(t) m^-(t)^p \Big) \Delta t \bigg]} \sum_{\{ \sigma_j^{\pm}(t) \}, \{ \psi_j^{\pm}(t) \}} \prod_{t,r} \delta \Big( Nm^r(t) - \sum_j \sigma_j^r(t) \Big) \prod_j \mathcal{B}_j,
\end{aligned}
\end{equation}
and express the $\delta$-functions as integrals of complex exponentials ($\delta(x) \propto \int \textrm{d}h \, e^{\pm ihx}$):
\begin{equation} \label{eq:Keldysh_top_line_simplification_v2}
\begin{aligned}
&\prod_{t,r} \delta \Big( Nm^r(t) - \sum_j \sigma_j^r(t) \Big) \\
&\qquad \propto \int \prod_{t,r} \textrm{d}h^r(t) \exp{\bigg[ -iN \sum_t \Big( h^+(t) m^+(t) - h^-(t) m^-(t) \Big) \Delta t \bigg]} \exp{\bigg[ i \sum_{j,t} \Big( h^+(t) \sigma_j^+(t) - h^-(t) \sigma_j^-(t) \Big) \Delta t \bigg]},
\end{aligned}
\end{equation}
where the omitted prefactors are unimportant.
Thus the generating functional becomes
\begin{equation} \label{eq:Keldysh_generating_functional_almost_path_integral}
\begin{aligned}
\mathcal{Z} &\propto \int \prod_{t,r} \textrm{d}m^r(t) \textrm{d}h^r(t) \exp{\bigg[ iN \sum_t \bigg( s(t) m^+(t)^p - s(t) m^-(t)^p - h^+(t) m^+(t) + h^-(t) m^-(t) \bigg) \Delta t \bigg]} \\
&\quad \cdot \sum_{\{ \sigma_j^{\pm}(t) \}, \{ \psi_j^{\pm}(t) \}} \exp{\bigg[ i \sum_{j,t} \big( h^+(t) \sigma_j^+(t) - h^-(t) \sigma_j^-(t) \big) \Delta t \bigg]} \prod_j \mathcal{B}_j.
\end{aligned}
\end{equation}
The sum over basis states now factors across $j$: the lower line of Eq.~\eqref{eq:Keldysh_generating_functional_almost_path_integral} can be written $\prod_j \overline{\mathcal{Z}}_j[h^+, h^-]$, where
\begin{equation} \label{eq:Keldysh_single_spin_generating_functional}
\overline{\mathcal{Z}}_j[h^+, h^-] \equiv \sum_{\{ \sigma^{\pm}(t) \}, \{ \psi^{\pm}(t) \}} \exp{\bigg[ i \sum_t \big( h^+(t) \sigma^+(t) - h^-(t) \sigma^-(t) \big) \Delta t \bigg]} \mathcal{B}_j.
\end{equation}
We thus have the desired path-integral representation of the generating functional:
\begin{equation} \label{eq:Keldysh_generating_functional_path_integral}
\mathcal{Z} \propto \int \prod_{t,r} \textrm{d}m^r(t) \textrm{d}h^r(t) \exp{\bigg[ iN \sum_t \bigg( s(t) m^+(t)^p - s(t) m^-(t)^p - h^+(t) m^+(t) + h^-(t) m^-(t) \bigg) \Delta t + \sum_j \log{\overline{\mathcal{Z}}_j[h^+, h^-]} \bigg]}.
\end{equation}

At large $N$, we can evaluate the path integral by saddle-point approximation (this is why our results only hold in the $N \rightarrow \infty$ limit).
Although the result must be $\mathcal{Z} = 1$, note that if we repeat this derivation for the average magnetization at time $t$, which can be written
\begin{equation} \label{eq:Keldysh_average_magnetization_definition}
\frac{1}{N} \sum_j \textrm{Tr} \big[ \rho(t) \hat{\sigma}_j^z \big] = \textrm{Tr} \Big( \mathcal{T} e^{-i \int_0^{\tau} \textrm{d}t H(t)} \Big) \rho(0) \Big( \mathcal{T} e^{-i \int_0^t \textrm{d}t' H(t')} \Big)^{\dagger} \bigg( \frac{1}{N} \sum_j \hat{\sigma}_j^z \bigg) \Big( \mathcal{T} e^{-i \int_t^{\tau} \textrm{d}t' H(t')} \Big)^{\dagger},
\end{equation}
then due to introducing the $\delta$-function in Eq.~\eqref{eq:Keldysh_top_line_simplification_v1}, the result is the same path integral as in Eq.~\eqref{eq:Keldysh_generating_functional_path_integral} with an additional factor of $m^-(t)$.
According to the saddle-point approximation, this integrates to simply the saddle-point value of $m^-(t)$.
Note that we can equally well relate the magnetization to $m^+(t)$ by starting from the expression $N^{-1} \textrm{Tr}[\hat{\sigma}_j^z \rho(t)]$ instead.
Either way, we determine the magnetization by locating the saddle point of Eq.~\eqref{eq:Keldysh_generating_functional_path_integral}.

Setting the derivatives of the exponent to zero gives the saddle-point equations
\begin{equation} \label{eq:Keldysh_saddle_point_equations_m}
h^+(t) = s(t) p m^+(t)^{p-1}, \qquad h^-(t) = s(t) p m^-(t)^{p-1},
\end{equation}
\begin{equation} \label{eq:Keldysh_saddle_point_equations_h}
m^+(t) = \frac{1}{iN \Delta t} \sum_j \frac{\partial \log{\overline{\mathcal{Z}}_j[h^+, h^-]}}{\partial h^+(t)}, \qquad m^-(t) = -\frac{1}{iN \Delta t} \sum_j \frac{\partial \log{\overline{\mathcal{Z}}_j[h^+, h^-]}}{\partial h^-(t)}.
\end{equation}
A consistent solution to these equations has $m^+(t) = m^-(t)$ and $h^+(t) = h^-(t)$.
This is clear in Eq.~\eqref{eq:Keldysh_saddle_point_equations_m}.
To see it in Eq.~\eqref{eq:Keldysh_saddle_point_equations_h}, first note that if we set $h^+(t) = h^-(t) = h(t)$, then $\overline{\mathcal{Z}}_j$ becomes the Trotterized generating functional for a \textit{single} spin and bath evolving under Hamiltonian $B_j(t) - h(t) \hat{\sigma}_j^z$: as noted above, $\mathcal{B}_j$ is the collection of terms corresponding to evolution under $B_j(t)$, and the additional exponential in Eq.~\eqref{eq:Keldysh_single_spin_generating_functional} is what results from adding a longitudinal field $h(t)$.
Furthermore, $\partial \log{\overline{\mathcal{Z}}_j}/\partial h^{\pm}(t)$ brings down a factor of $\pm i \sigma^{\pm}(t) \Delta t$ and thus amounts to the Trotterized expression for $\pm i \langle \hat{\sigma}_j^z \rangle \Delta t$ --- both right-hand sides in Eq.~\eqref{eq:Keldysh_saddle_point_equations_h} are then $N^{-1} \sum_j \langle \hat{\sigma}_j^z \rangle$, and it is indeed consistent to set $m^+(t) = m^-(t) = m(t)$.

To summarize, the saddle-point equations that determine the magnetization $m(t)$ amount to
\begin{equation} \label{eq:Keldysh_simple_saddle_point_equations}
h(t) = s(t) p m(t)^{p-1}, \qquad m(t) = \frac{1}{N} \sum_j \big< \hat{\sigma}_j^z(t) \big>,
\end{equation}
where, importantly, the expectation values are for \textit{uncoupled} spins (each with its own bath) evolving under Hamiltonian $B_j(t) - h(t) \hat{\sigma}_j^z$.
Using Eq.~\eqref{eq:full_Hamiltonian_decomposition_B} for $B_j(t)$, we can write
\begin{equation} \label{eq:Keldysh_effective_spin_bath_Hamiltonian}
B_j(t) - h(t) \hat{\sigma}_j^z = \overline{H}_j(t) + \sum_k \omega_k \hat{b}_{jk}^{\dagger} \hat{b}_{jk} + g \sum_k \kappa_k \hat{\sigma}_j^z \big( \hat{b}_{jk} + \hat{b}_{jk}^{\dagger} \big),
\end{equation}
where the effective Hamiltonian for spin $j$ is
\begin{equation} \label{eq:Keldysh_effective_spin_Hamiltonian}
\overline{H}_j(t) = -\Big[ h(t) + \big( 1 - s(t) \big) \big( 1 - \lambda(t) \big) \epsilon_j \Big] \hat{\sigma}_j^z - \big( 1 - s(t) \big) \lambda(t) \hat{\sigma}_j^x,
\end{equation}
in agreement with Eq.~\eqref{eq:effective H_j} from the main text.

Even though $h(t)$ is itself a function of the magnetization, Eq.~\eqref{eq:Keldysh_effective_spin_bath_Hamiltonian} still amounts to a spin subject to time-dependent longitudinal and transverse fields while coupled to a bath.
We can thus follow Ref.~\cite{quantum_markovian_master_equation} to derive a time-dependent master equation for the \textit{single-spin} density matrix $\rho_j(t)$, such that $\langle \hat{\sigma}_j^z \rangle = \textrm{Tr}[\rho_j(t) \hat{\sigma}_j^z]$.
Since we do not modify the analysis at all, we give only the final result and refer to Ref.~\cite{quantum_markovian_master_equation} for details, although it is important to note that the derivation only holds at weak coupling and in the adiabatic limit.
Precise conditions are given in Ref.~\cite{quantum_markovian_master_equation}, but since here we use values comparable to 1 for all other parameters, it is sufficient to take $g \ll 1$ and $\tau \gg 1$ (note that all parameters are dimensionless since we have set the energy scale in the $p$-spin model to 1).

Ultimately, the saddle-point equations become
\begin{equation} \label{eq:Keldysh_open_saddle_point_equations}
h(t) = s(t) p m(t)^{p-1}, \qquad m(t) = \frac{1}{N} \sum_j \textrm{Tr} \big[ \rho_j(t) \hat{\sigma}_j^z \big],
\end{equation}
where $\rho_j(t)$ is governed by the equation
\begin{equation} \label{eq:effective_single_spin_Lindbladian}
\partial_t \rho_j(t) = -i \big[ \overline{H}_j(t) + H_{LS}(t), \rho_j(t) \big] + \mathcal{D} \big[ \rho_j(t) \big],
\end{equation}
and all quantities are as follows.
The effective Hamiltonian $\overline{H}_j(t)$ is that in Eq.~\eqref{eq:Keldysh_effective_spin_Hamiltonian}, the Lamb-shift Hamiltonian is
\begin{equation} \label{eq:single_spin_Lamb_shift}
H_{LS}(t) = \sum_{\omega} S(\omega) L_{\omega}(t)^{\dagger} L_{\omega}(t),
\end{equation}
and the dissipator is
\begin{equation} \label{eq:single_spin_dissipator}
\mathcal{D} \big[ \rho_j(t) \big] = \sum_{\omega} \gamma(\omega) \Big( L_{\omega}(t) \rho_j(t) L_{\omega}(t)^{\dagger} - \frac{1}{2} L_{\omega}(t)^{\dagger} L_{\omega}(t) \rho_j(t) - \frac{1}{2} \rho_j(t) L_{\omega}(t)^{\dagger} L_{\omega}(t) \Big).
\end{equation}
The sums over $\omega$ run over all differences between eigenvalues of $\overline{H}_j(t)$.
The relaxation rate $\gamma(\omega)$ is the Fourier transform of the bath correlation function, and $S(\omega)$ is its Hilbert transform, but it is standard to simply choose an expression for $\gamma(\omega)$ rather than derive it --- here we use
\begin{equation} \label{eq:Ohmic_bath_relaxation_rate}
\gamma(\omega) = 2\pi \eta \frac{\omega e^{-|\omega|/\omega_c}}{1 - e^{-\beta \omega}}, \qquad S(\omega) = \mathcal{P} \int \frac{\textrm{d}\omega'}{2\pi} \frac{\gamma(\omega')}{\omega - \omega'},
\end{equation}
where $\mathcal{P}$ denotes the principal part and we set $\eta$ (proportional to $g^2$) to $10^{-3}$.
Lastly, the jump operator $L_{\omega}(t)$ is
\begin{equation} \label{eq:single_spin_jump_operator}
L_{\omega}(t) = \sum_{a,b} \delta_{\varepsilon_b(t) - \varepsilon_a(t), \omega} \big| \varepsilon_a(t) \big> \big< \varepsilon_a(t) \big| \hat{\sigma}_j^z \big| \varepsilon_b(t) \big> \big< \varepsilon_b(t) \big|,
\end{equation}
where $a$ and $b$ run over the eigenstates of $\overline{H}_j(t)$.

Since $\overline{H}_j(t)$ is a $2 \times 2$ matrix, we can be more explicit.
In terms of the overall longitudinal and transverse fields,
\begin{equation} \label{eq:total_single_spin_fields}
\nu(t) \equiv h(t) + (1 - s(t))(1 - \lambda(t)) \epsilon_j, \qquad \Gamma(t) \equiv (1 - s(t)) \lambda(t),
\end{equation}
the eigenvalues and eigenvectors of $\overline{H}_j(t)$ are
\begin{equation} \label{eq:effective_single_spin_eigenstates}
\varepsilon_{\pm}(t) = \pm \sqrt{\nu(t)^2 + \Gamma(t)^2}, \qquad \big| \varepsilon_{\pm}(t) \big> = \frac{1}{\sqrt{(\varepsilon_{\pm}(t) + \nu(t))^2 + \Gamma(t)^2}} \begin{pmatrix} \Gamma(t) \\ -(\varepsilon_{\pm}(t) + \nu(t)) \end{pmatrix}.
\end{equation}
Thus the three frequencies being summed over in Eqs.~\eqref{eq:single_spin_Lamb_shift} and~\eqref{eq:single_spin_dissipator} are
\begin{equation} \label{eq:single_spin_Bohr_frequencies}
\omega \in \big\{ 0, \omega_+, \omega_- \big\}, \qquad \omega_{\pm} \equiv \pm 2 \sqrt{\nu(t)^2 + \Gamma(t)^2},
\end{equation}
with corresponding jump operators
\begin{equation} \label{eq:single_spin_jump_operator_0}
L_0(t) = \frac{\nu(t)}{\nu(t)^2 + \Gamma(t)^2} \begin{pmatrix} \nu(t) & \Gamma(t) \\ \Gamma(t) & -\nu(t) \end{pmatrix},
\end{equation}
\begin{equation} \label{eq:single_spin_jump_operator_pm}
L_{\pm}(t) = \frac{\Gamma(t)}{2 [\nu(t)^2 + \Gamma(t)^2]} \begin{pmatrix} \Gamma(t) & \mp \sqrt{\nu(t)^2 + \Gamma(t)^2} - \nu(t) \\ \pm \sqrt{\nu(t)^2 + \Gamma(t)^2} - \nu(t) & -\Gamma(t) \end{pmatrix}.
\end{equation}

Solving the saddle-point equations numerically is straightforward.
Since $\epsilon_j$ only takes the values $1$ and $-1$, and $\rho_j$ begins in the state $| \epsilon_j \rangle \langle \epsilon_j |$ for the ARA protocol, the solution of Eq.~\eqref{eq:effective_single_spin_Lindbladian} is identical for all spins having the same value of $\epsilon_j$.
We only need to determine the two density matrices $\rho_u(t)$ and $\rho_d(t)$, the states for spins having $\epsilon_j = 1$ and $\epsilon_j = -1$ respectively.
Recalling that $c$ is the fraction of spins with $\epsilon_j = 1$, the self-consistency condition becomes $m(t) = c \textrm{Tr}[\rho_u(t) \hat{\sigma}^z] + (1-c) \textrm{Tr}[\rho_d(t) \hat{\sigma}^z]$, and note that $\rho_{u(d)}(t)$ depends only on the magnetization at previous times.
Thus the algorithm for determining $m(t)$ is as follows (we use $\Delta t = 0.1$ in the main text):
\begin{enumerate}
\item Begin with a spin in state $\rho_u(0) = | \uparrow \; \rangle \langle \; \uparrow |$, and a spin in state $\rho_d(0) = | \downarrow \; \rangle \langle \; \downarrow |$.
\item Do the following in order, setting $t = 0$:
\begin{itemize}
\item Calculate
\begin{equation} \label{eq:algorithm_magnetization}
m(t) = c \textrm{Tr} \big[ \rho_u(t) \hat{\sigma}^z \big] + (1 - c) \textrm{Tr} \big[ \rho_d(t) \hat{\sigma}^z \big].
\end{equation}
\item Set
\begin{equation} \label{eq:algorithm_field}
h(t) = s(t) p m(t)^{p-1}.
\end{equation}
\item Evolve $\rho_u(t)$ for time $\Delta t$ under the master equation in Eq.~\eqref{eq:effective_single_spin_Lindbladian}, where the Hamiltonian from which all quantities are derived (Eqs.~\eqref{eq:single_spin_Lamb_shift} through~\eqref{eq:single_spin_jump_operator_pm}) is
\begin{equation} \label{eq:algorithm_Hamiltonian_up}
H_u(t) = -\Big[ h(t) + \big( 1 - s(t) \big) \big( 1 - \lambda(t) \big) \Big] \hat{\sigma}_j^z - \big( 1 - s(t) \big) \lambda(t) \hat{\sigma}_j^x.
\end{equation}
Similarly evolve $\rho_d(t)$ for time $\Delta t$ under the same type of master equation but using Hamiltonian
\begin{equation} \label{eq:Keldysh_self_consistent_Hamiltonian_down}
H_d(t) = -\Big[ h(t) - \big( 1 - s(t) \big) \big( 1 - \lambda(t) \big) \Big] \hat{\sigma}_j^z - \big( 1 - s(t) \big) \lambda(t) \hat{\sigma}_j^x.
\end{equation}
This gives the states $\rho_u(t + \Delta t)$ and $\rho_d(t + \Delta t)$.
\end{itemize}
\item Now that $\rho_u(\Delta t)$ and $\rho_d(\Delta t)$ have been determined, repeat step 2 for $t = \Delta t$ to obtain $\rho_u(2 \Delta t)$ and $\rho_d(2 \Delta t)$, then repeat again to obtain $\rho_u(3 \Delta t)$ and $\rho_d(3 \Delta t)$, and so on.
\end{enumerate}

\end{widetext}

\end{document}